\documentclass[prl,aps,superscriptaddress,floatfix,showpacs,preprintnumbers,amsmath,amssymb,amsfonts,lengthcheck]{revtex4-2}

\usepackage{graphicx}
\usepackage{bm}
\usepackage{braket}
\usepackage{xcolor}
\usepackage{dutchcal}
\usepackage{microtype}
\usepackage[colorlinks,allcolors=blue]{hyperref}

\def\equationautorefname#1#2\null{Eq.#1(#2\null)}

\renewcommand{\vec}[1]{\mathbf{#1}}
\newcommand{\tens}[1]{\mbox{\textsf{\textbf{#1}}}}
\newcommand{\dif}{\!\! \mathrm{d}}
\newcommand{\mi}{\textrm{i}} 
\newcommand{\me}{\mathrm{e}}
\newcommand{\kparb}{\overline{\vec{k}}_\parallel}

\newcommand{\ethz}{Institute for Theoretical Physics, ETH Zürich, CH-8093 Zürich, Switzerland}
\newcommand{\madridone}{Departamento de Física Teórica de la Materia Condensada, Universidad Autónoma de Madrid, E-28049 Madrid, Spain}
\newcommand{\madridtwo}{Condensed Matter Physics Center (IFIMAC), Universidad Autónoma de Madrid, E-28049 Madrid, Spain}

\begin{document}

\author{Frieder Lindel}
%\email{flindel@ethz.ch}
\affiliation{\madridone}
\affiliation{\madridtwo}
\affiliation{\ethz}

\author{Carlos J. Sánchez Martínez}
\affiliation{\madridone}
\affiliation{\madridtwo}

\author{Johannes Feist}
\email{johannes.feist@uam.es}
\affiliation{\madridone}
\affiliation{\madridtwo}

\author{Francisco J. García-Vidal}
\email{fj.garcia@uam.es}
\affiliation{\madridone}
\affiliation{\madridtwo}

\title{Close encounters between periodic light and periodic arrays of quantum emitters}

\date{\today}

\begin{abstract}
We introduce crystal polaritons, hybrid excitations formed when the collective
excitations of a periodic quantum-emitter array strongly couple to the resonant Bloch modes
of a metasurface. This realizes a cavity-QED platform in which periodic light and periodic
matter are treated on the same footing, allowing strong collective light-matter coupling in an
extended, lossy, and dispersive nanophotonic structure. To describe this regime, we develop a
reciprocal-space few-mode quantization based on macroscopic quantum electrodynamics,
which maps the metasurface resonances seen by the emitter array onto a cavity-QED
Hamiltonian at each in-plane momentum. We show that both plasmonic surface-lattice
resonances and dielectric bound states in the continuum can enter the strong-coupling regime
with a single emitter per unit cell. As a consequence of the resonant nonlinearities of the
resulting crystal polaritons, the platform enables quantum light generation with efficiencies 
orders of magnitude higher than those achieved in conventional nonlinear metasurfaces.
\end{abstract}

\maketitle

Confining photons in cavities can lead to strong coupling of single or few photonic modes to quantum emitters. This is at the heart of cavity quantum electrodynamics (QED), with applications in the generation of non-classical states of light \cite{haroche2006exploring}, quantum sensors \cite{degen2017quantum}, or in quantum information science \cite{kimble2008quantum}. However, traditional wavelength-sized cavities are restricted by the diffraction limit. While plasmonic nanoparticles overcome this via subwavelength confinement \cite{chang2018colloquium,fregoni2022theoretical,gonzalez2024light}, they suffer from high losses that limit many quantum optical applications.

\begin{figure}%[t]
    \centering
    \includegraphics[width=1\linewidth]{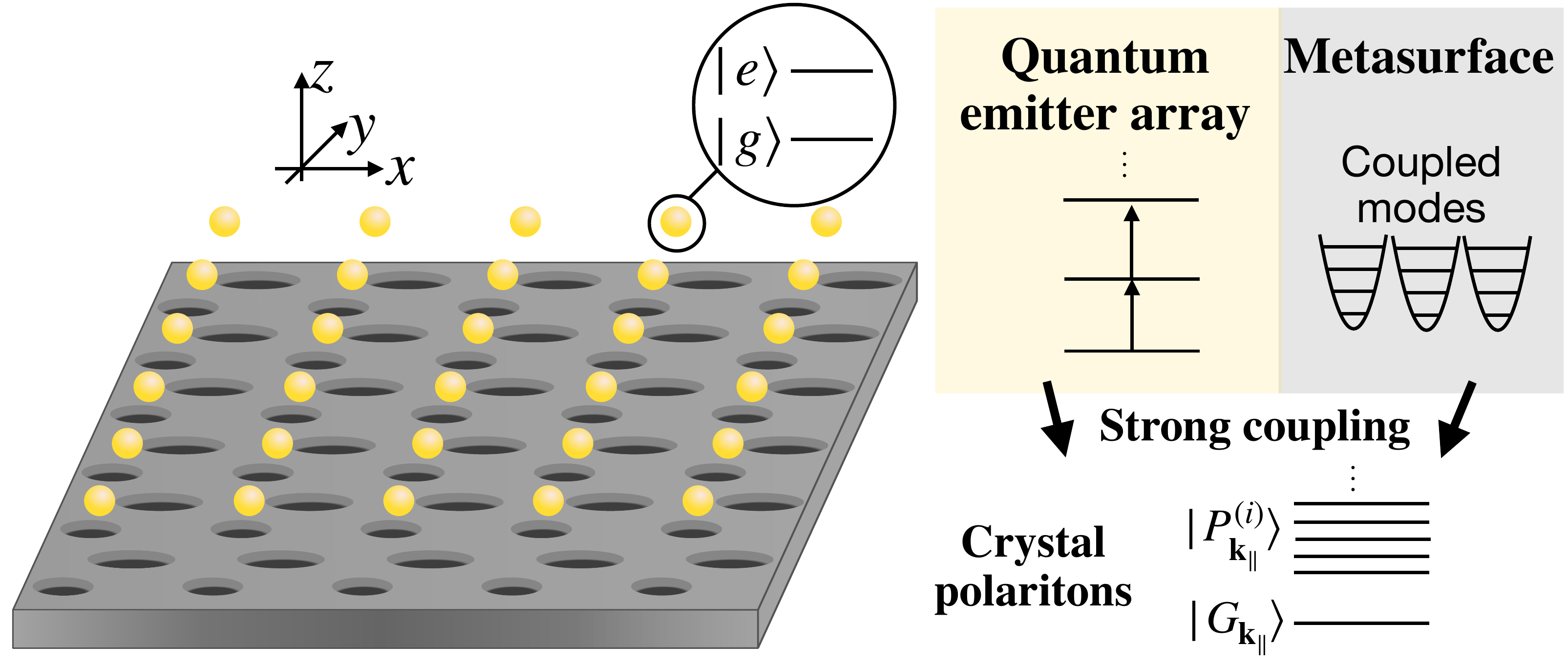}
    \caption{\textit{Concept of a polaritonic quantum metasurface hosting crystal polaritons.} A metasurface confines the light field in its vicinity that is strongly coupled to collective excitations of a quantum emitter array. Strong coupling of the quantized modes of the metasurface to the emitter array excitations leads to the formation of crystal polaritons.}
    \label{fig:setup}
\end{figure}

An alternative is provided by two-dimensional periodically structured surfaces (metasurfaces). Because their resonant modes extend over many unit cells, metasurfaces can combine subwavelength field confinement with mitigated losses \cite{zou2004silver}. They offer many exciting design possibilities to, e.g., support flat bands \cite{mukherjee2015observation}, topologically protected states \cite{lu2016symmetry,ni2023topological} and lasing \cite{bahari2017nonreciprocal,bandres2018topological,han2020lasing}, or bound states in the continuum \cite{fan2002analysis,plotnik2011experimental,hsu2013observation,hsu2016bound}. Consequently, they have become a well-established tool for manipulating classical \cite{shalaev2007optical,joannopoulos1997photonic} and quantum states of light \cite{santiago2022resonant,solntsev2021metasurfaces}, and have been used for strong light-matter coupling to dense molecular ensembles \cite{doi:10.1021/nl4035219,do2024room} leading to polariton lasing \cite{ramezani2016plasmon,vakevainen2020sub}, polariton BECs \cite{hakala2018bose}, and polaritonic chemistry \cite{verdelli2024polaritonic}.  

Arrays of quantum emitters in free space---such as cold atoms or molecules---have emerged as platforms for cooperative optics with applications in classical optics \cite{shahmoon2017cooperative,perczel2017photonic,rui2020subradiant}, nonlinear quantum optics \cite{bettles2020quantum,moreno2021quantum,zhang2022photon,pedersen2023quantum,pedersen2024green,dura2025quantum}, or quantum information processing \cite{asenjo2017exponential,bluvstein2024logical} and quantum simulation \cite{morgado2021quantum}. 

Despite the versatility of these two periodic platforms, their combination has largely been restricted to the weak-coupling regime \cite{chang2018colloquium,reitz2022cooperative}, and previous \emph{ab-initio} frameworks that can accurately determine the quantized resonant modes of metasurfaces to which nearby emitters strongly couple are limited to the coupled-dipole approximation and linear response \cite{reitz2025quantum}.

 \begin{figure}
    \centering
    \includegraphics[width=1\linewidth]{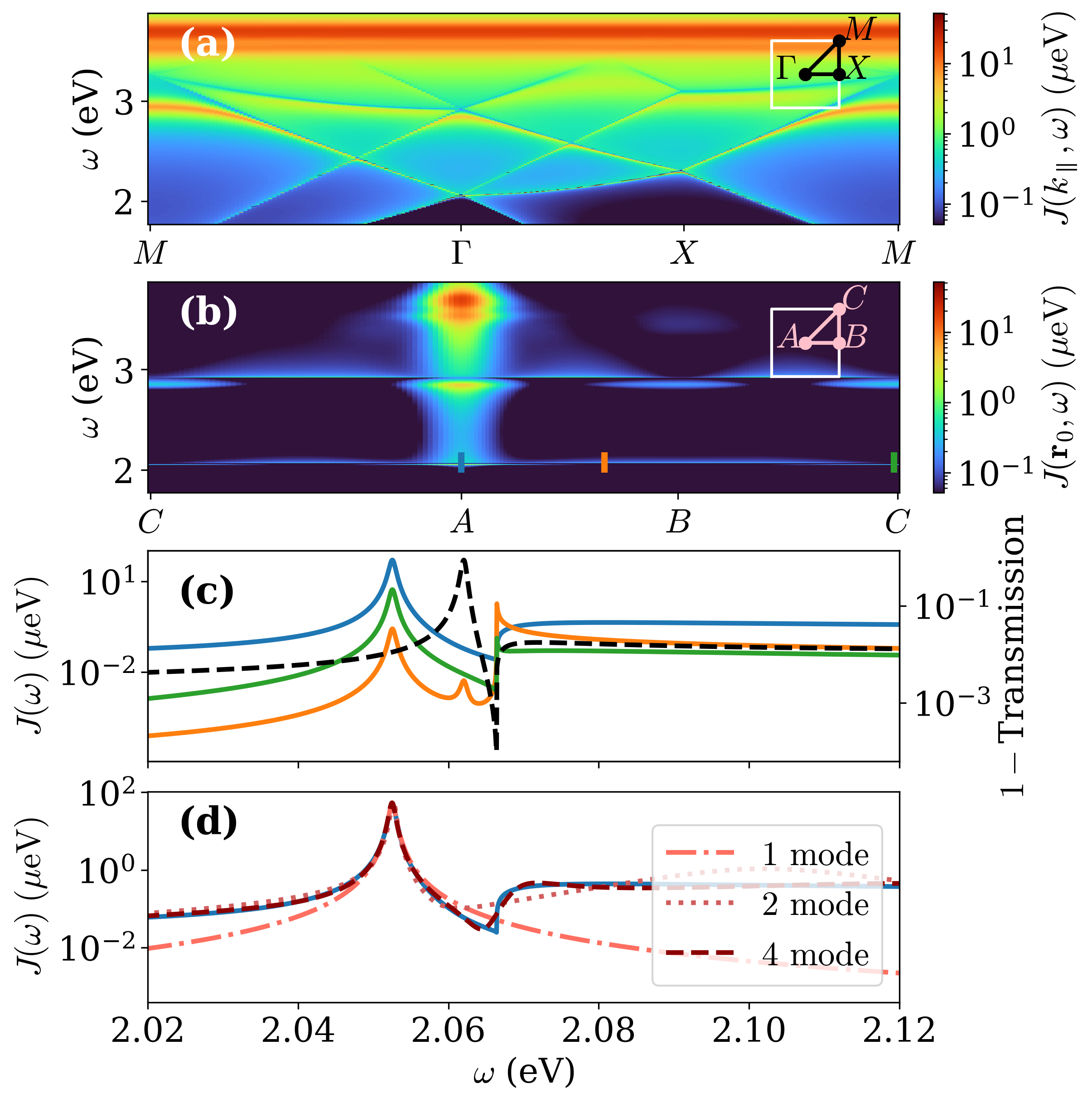}
    \caption{\textit{Optical environment seen by a collective emitter-array close to a square array of silver nanospheres.} (a) Reciprocal space spectral density $J$ as a function of frequency and in-plane wave-vector along a path in reciprocal space as indicated by the inset for an emitter array positioned in the center of the unit cell [$\vec{r}_\parallel = (0,0)$] and $10\,$nm above the nanospheres with dipole moment $d= 10\,$D and orientation $\vec{n} = \vec{e}_z$. (b) $J$ at the $\Gamma$ point ($\vec{k}_\parallel = \vec{0}$) for different emitter positions indicated by the colored rectangles in (b) ($\vec{r}_\parallel = (0,0)$, $\vec{r}_\parallel = (200,0)\mathrm{nm}$, $\vec{r}_\parallel = (300,300)\mathrm{nm}$). (c) Cut through $J$ in (b) at three different positions indicated by vertical lines in (b) (colored lines) and the transmission through the array (black dashed line). (d) $J$ for an emitter at $A$ and the $\Gamma$ point shown in (c) together with the fitted model spectral density $J_\mathrm{mod}$ [Eq.~\eqref{eq:SpectralDensityMod}] with different numbers of modes.}
    \label{fig:Metallic_array}
\end{figure}

The regime we target here fundamentally differs from standard cavity QED and strong coupling of dense molecular ensembles to resonances supported by periodic nanostructures \cite{doi:10.1021/nl4035219,do2024room, ramezani2016plasmon,vakevainen2020sub,hakala2018bose,verdelli2024polaritonic}. In a conventional cavity, a small number of \emph{localized} photonic modes couple to localized matter excitations. By contrast, the relevant optical resonances in a metasurface are extended Bloch modes. As a difference also to dense ensembles of molecules, which behave essentially as classical harmonic oscillators, the discrete quantum emitters show much higher nonlinearities and therefore bear the potential to enter a nonlinear cavity QED regime. If a quantum-emitter array is made commensurable with the metasurface [see \autoref{fig:setup}], this opens the possibility for coupling of collective spin-wave excitations of the emitter array with metasurface resonances at well defined Bloch momentum. The central question is then whether these two extended, periodic degrees of freedom--photonic and material--can hybridize strongly. 

In this Letter, we show that this is indeed possible. By identifying the coupled, lossy photonic modes of general metasurfaces via macroscopic QED, we obtain a cavity QED-type Hamiltonian that gives full access to the quantum dynamics in the strong-coupling regime. We find that strong coupling is achievable with just a single emitter per unit cell, coupling to either bound states in the continuum of a dielectric array of nanoparticles or to surface lattice resonances of a metallic array of nanoparticles. This results in the hybridization of emitter spin-waves and metasurface resonances into \emph{crystal polaritons}. Furthermore, we show that laser driving of these crystal polaritons yields resonant quantum light generation with efficiencies many orders of magnitude higher than state-of-the-art nonlinear metasurfaces \cite{liu2019high,koshelev2019nonlinear,yang2015nonlinear,grinblat2021nonlinear}, reaching a strong nonlinear response with just one crystal polariton per $10.7\, \mu\mathrm{m}^2$.

\paragraph{Collective light–matter coupling in reciprocal space.}

The essential simplification of the commensurable quantum metasurface [see Fig.~\ref{fig:setup}] is that both subsystems share the same periodicity. The electromagnetic resonances of the metasurface are Bloch modes, and the collective excitations of the emitter array are spin waves carrying the same in-plane momentum. Strong coupling therefore does not occur between a localized emitter and a localized cavity mode, but between two extended Bloch-periodic degrees of freedom with the same Bloch momentum. The first task is to identify, for each in-plane momentum $\mathbf{k}_\parallel$ inside the Brillouin zone, which metasurface resonances are available to hybridize with the corresponding emitter-array excitation. As we show in the following, this information is contained in a reciprocal-space spectral density. It plays the same role as the local spectral density in ordinary cavity QED \cite{breuer2002theory,menczel2024non}, but resolved in Bloch momentum and in the emitter positions within the unit cell. It determines the complex resonant frequencies of the electromagnetic environment seen by the spin wave and the strength with which this spin wave couples to them.

We characterize the emitters by their dipole operator $\hat{d}^{(h)}_i = d_i (\hat{\sigma}^{(h)}_i + \hat{\sigma}^{(h) \dagger}_i)$ and orientation $\vec{n}^{(h)}$, as well as their position $\vec{r}^{(h)}_i = \vec{r}_0^{(h)} + \vec{R}_i$, with $\vec{r}_0^{(h)}$ denoting the position of the $h$th dipole in the reference unit cell and $ \left\lbrace\vec{R}_i\right\rbrace$ being the set of lattice vectors connecting the different unit cells. As we derive in detail in the Supplemental Material (SM) \cite{supplemental_material}, the Bloch-periodic spin-wave excitations of such emitter arrays $
    \hat{d}^{(h)}_{\vec{k}_\parallel}=   \sum_{i} \hat{d}_{i}^{(h)} \me^{-\mi \vec{k}_\parallel \cdot \vec{r}_{i}^{(h)}}$ couple to the continuum of field modes via the reciprocal-space spectral density of the metasurface
\begin{multline} \label{eq:spectraldensity}
J_{h,h^\prime}(\vec{k}_\parallel, \omega) = \frac{ \mu_0 \omega^2 }{\pi \hbar} d_h d_{h^\prime}  \me^{-\mi \vec{k}_\parallel (r_0^{(h)}-r_0^{(h^\prime)})} \\ \times
\vec{n}^{(h)} \cdot  \boldsymbol{\mathcal{G}}_\mathrm{AH}(\vec{k}_\parallel, \vec{r}_0^{(h)},\vec{r}_0^{(h^\prime)} ,\omega)\cdot \vec{n}^{(h^\prime)} .
\end{multline} 
Here, $\boldsymbol{\mathcal{G}}_\mathrm{AH}$ is the anti-Hermitian part of the Bloch-periodic Green tensor $ \boldsymbol{\mathcal{G}} (\vec{k}_\parallel, \vec{r},\vec{r}^\prime ,\omega) =  \sum_{i} \me^{\mi \vec{k}_\parallel \cdot \vec{R}_i}  \tens{G} ( \vec{r},\vec{r}^\prime+\vec{R}_i ,\omega) $ \cite{jorgenson2002efficient,linton2010lattice} and dispersion and absorption was accounted for via macroscopic quantum electrodynamics \cite{scheel_macroscopic_2008}.

We have thus identified $J_{h,h^\prime}(\vec{k}_\parallel, \omega) $ as the key quantity that fully determines the optical environment of the metasurface seen by the emitter array and thus characterizes the extended resonances of metasurfaces. To illustrate this quantity, we consider a square array of silver nanospheres with radius $R = 50\,$nm and lattice constant $a= 600\,$nm that is coupled to an array of two-level emitters with a single emitter per unit cell located at different positions within the unit cell. The dipole moment of the two-level emitters is given by $d = 10\,$D and the response of the silver spheres is described by a Drude permittivity as in Refs.~\cite{delga2014quantum,palik1998handbook}. 

As shown in \autoref{fig:Metallic_array}, the maxima of the reciprocal-space spectral density $J \equiv J_{1,1}(\vec{k}_\parallel, \omega) $ follow the usual dispersion of the surface lattice resonances of the plasmonic array, resulting from the coupling of diffractive orders (Rayleigh anomalies) with the plasmonic resonance of the nanoparticles (here at $3.5\,$eV) \cite{cherqui2019plasmonic}. Nevertheless, $J$ differs qualitatively from far-field observables such as reflectance, transmittance or extinction of the array, see \autoref{fig:Metallic_array}(c). In contrast to these quantities, $J$ depends on where the emitter sits inside the unit cell and on its dipole orientation, because it describes the collective near-field optical environment experienced by the matter excitation rather than the response of the bare metasurface to an external probe [see \autoref{fig:Metallic_array}(b)].

 \begin{figure}%[b]
    \centering
    \includegraphics[width=1\linewidth]{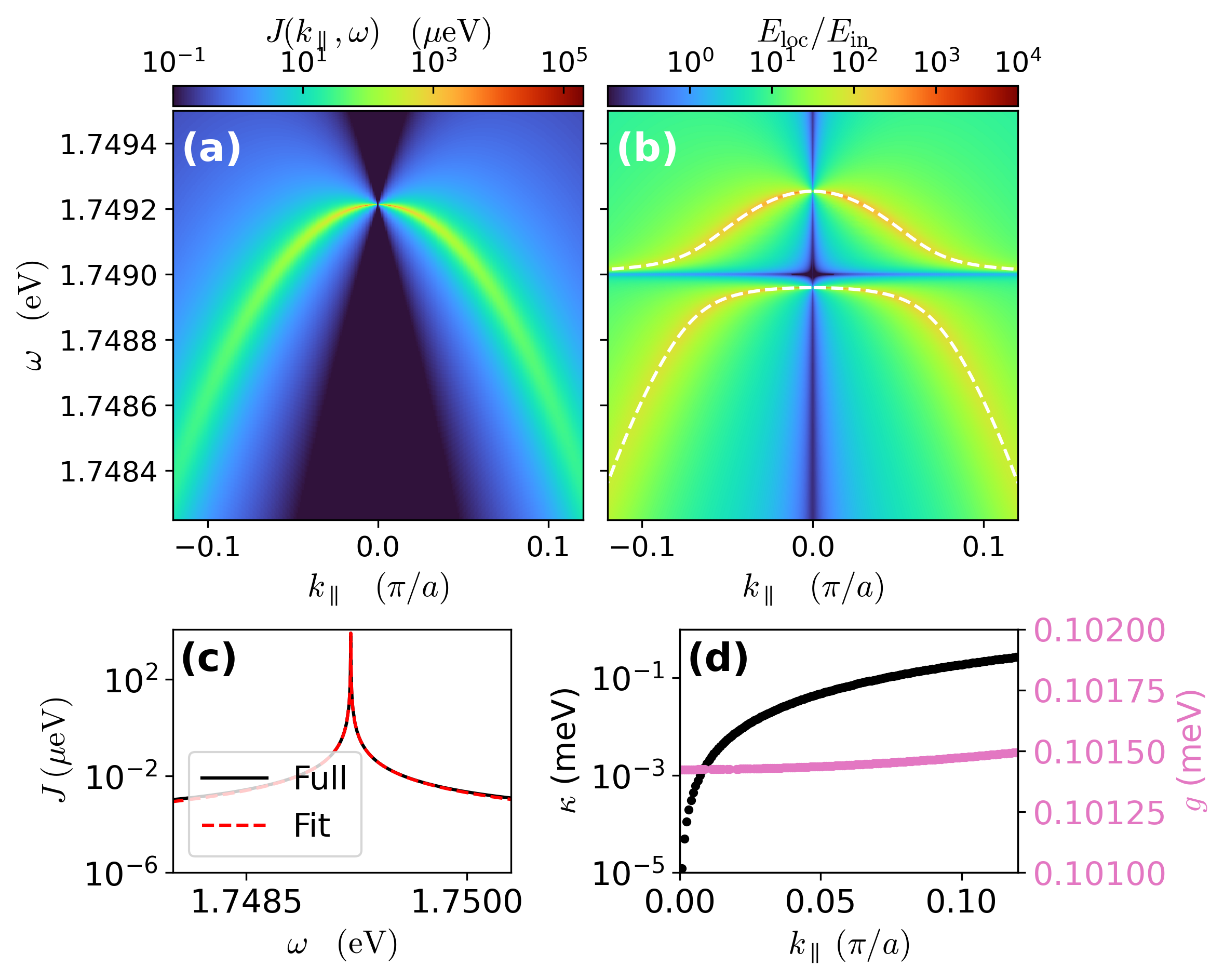}
    \caption{\textit{Crystal-polariton formation with a bound state in the continuum.} (a) Dispersion of the reciprocal-space spectral density $J$ for a square array of dielectric (silicon) nanospheres with radius $R = 100\,$nm, refractive index $n=3.5$, lattice constant $a=400\,$nm, $\vec{k}_\parallel = k_\parallel \vec{e}_y$, and the emitter array is positioned at $(0,105\,\mathrm{nm},0)$ in the unit cell with frequency $\omega_E = 1.749\,$eV, dipole moment $d= 10\,$D, and orientation $\vec{n} = \vec{e}_x$. (b) Local field strength $E_\mathrm{loc}$ at the position of the emitter as a function of the frequency $\omega_L$ and in-plane momentum $\vec{k}_\mathrm{L} = k_\mathrm{L} \vec{e}_y$ of the plane-wave laser driving with amplitude $E_\mathrm{in}$ and polarization $\vec{e}_x$. The dispersion of the crystal polaritons is shown by the dashed lines. (c) Exemplary single-mode fit of $J$ at $k_\parallel =6.4\times 10^{-3} \pi/a$. (d) Coupling strength $g$ and decay rate $\kappa$ of the bound state in the continuum obtained from fitting a single-mode model to $J$ at different in-plane momenta.}
    \label{fig:BIC}
\end{figure}

\paragraph{From the spectral density to a cavity-QED Hamiltonian}

The reciprocal-space spectral density identifies the optical resonances that the emitter-array
spin waves can see, which we now map to a set of coupled lossy modes with spectral density identical to the {\it ab initio} result via a recently developed few-mode quantization scheme \cite{tamascelli2018nonperturbative,lentrodt2020ab,medina_few-mode_2020,sanchez2022few,sanchez2024mixed,lednev2025spatially}. This step maps our setup to a typical cavity-QED Hamiltonian, with photonic Bloch modes and material spin waves appearing in equal terms.

The resulting few-mode Hamiltonian reads ($\hbar=1$):
\begin{multline}   \label{eq:few_mode_hamiltonian}
 \hat{H}_{\vec{k}_\parallel} =  \hat{H}_{e,\vec{k}_\parallel} + \sum_{i,j} \omega_{\vec{k}_\parallel,ij}  \hat{a}_{\vec{k}_\parallel,i}^\dagger \hat{a}_{\vec{k}_\parallel,j} \\  +   \sum_{i,h} ( g_{ \vec{k}_\parallel,ih} \hat{d}_{\vec{k}_\parallel}^{(h)\dagger}    \hat{a}_{\vec{k}_\parallel,i} + \mathrm{h.c.}) .
\end{multline}
Here, $\hat{H}_{e,\vec{k}_\parallel}$ is the bare emitter Hamiltonian. The operators $\hat a_{\mathbf{k}\parallel,i}$ describe the effective lossy photonic modes of the metasurface, while $\hat
d^{(h)}_{\mathbf{k}\parallel}$ describes the spin-wave excitations of the emitter array. Losses
are included through Lindblad terms with rates $\kappa_{\mathbf{k}\parallel,i}$, see SM \cite{supplemental_material}. The key point is that Eq.~\eqref{eq:few_mode_hamiltonian} is a cavity-QED Hamiltonian for an extended periodic platform: the coupling conserves Bloch momentum and hybridizes photonic and material excitations at the same $\mathbf{k}_\parallel$.

The parameters of Hamiltonian \eqref{eq:few_mode_hamiltonian} are not phenomenological. They are obtained by requiring the effective model to reproduce the ab initio reciprocal-space spectral density, $\tens{J}_{\rm
mod}(\mathbf{k}_\parallel,\omega )  =\tens{J}(\mathbf{k}_\parallel,\omega)$, with
\begin{align} \label{eq:SpectralDensityMod}
\tens{J}_\mathrm{mod}(\vec{k}_\parallel,\omega) = \frac{1}{\pi}  \tens{g}_{\vec{k}_\parallel }  \cdot \mathrm{Im} \left[ \frac{1}{\widetilde{\tens{H}}_{\vec{k}_\parallel} - \omega}  \right]\cdot  \tens{g}_{\vec{k}_\parallel }^\dagger ,
\end{align}
and $\widetilde{H}_{\vec{k}_\parallel,ij} = \omega_{\vec{k}_\parallel,ij} -\mi \kappa_{\vec{k}_\parallel,i} \delta_{ij}/2 $. This can be achieved by fitting the parameters of the few-mode model, namely the mode frequencies $\omega_{\vec{k}_\parallel,ii}$, the mode couplings $\omega_{\vec{k}_\parallel,ij}$, and the decay rates $\kappa_{\vec{k}_\parallel,i}$, as well as the light-matter coupling constants $g_{\vec{k}_\parallel,ih}$. %Thus, the resulting quantized modes are precisely those relevant for the emitter array, rather than modes inferred only from a far-field spectrum of the bare metasurface.
 
We apply this procedure to a representative SLR in the plasmonic array considered above; see \autoref{fig:Metallic_array}. Even when the far-field spectrum suggests that SLRs are single-mode resonances, we find that a naive single-mode picture of SLRs can fail:  the SLR line shape in the reciprocal-space spectral density is asymmetric, in contrast to the symmetric Lorentzian expected for a single mode [cf. Eq.~\eqref{eq:SpectralDensityMod}] \cite{lentrodt2023certifying}. Therefore, it cannot be reproduced by one Lorentzian mode. Physically, the asymmetry---and hence the multimode nature---can be attributed to the inherent asymmetry of the Rayleigh anomalies, which our few-mode model is able to fully capture, as seen in \autoref{fig:Metallic_array}(d).

\paragraph{Crystal polaritons in a quantum metasurface.}

We now use the developed framework to demonstrate our main physical result: a commensurable
emitter array and a metasurface can enter a collective strong-coupling regime in which the eigenmodes are hybrid light–matter
Bloch excitations. All elements of this experimental setup can be achieved with state-of-the-art devices. The main experimental challenge is to place the emitters in an array with fixed relative position to the nanoparticle. Promising platforms to do so are carbon-based color centers in hexagonal boron nitride films \cite{do2024room,tran2016quantum}, atoms trapped by Casimir--Polder forces of the metasurface \cite{gullans2012nanoplasmonic,gonzalez2015subwavelength}, or molecules placed via DNA origami \cite{kershner2009placement,kuzyk2018dna}.

For the plasmonic array considered above, we obtain the cavity QED parameters of the quantized surface lattice resonances from the fit of the few-mode model. We find that with a single emitter per unit cell that has a dipole moment of $10\,$D, the setup is close but does not quite reach the strong-coupling regime. As discussed in the SM \cite{supplemental_material}, the strong-coupling regime can be reached using slightly larger lattice constants $a>615\,$nm, however, requiring record high $Q$-factors \cite{bin2021ultra} for the SLRs.

Much higher quality factors than those of SLRs in plasmonic systems can be achieved for bound states in the continuum (BICs) in dielectric nanoparticle arrays \cite{liu2019high}. Therefore, we next consider a square array of dielectric (silicon) nanospheres and focus on a BIC originating from the magnetic dipole resonance of the spheres \cite{abujetas2021near}, see \autoref{fig:BIC}(a). We identify this BIC as an approximately Lorentzian resonance in the reciprocal-space spectral density, whose width narrows with decreasing in-plane wavevector, compare \autoref{fig:BIC}(c,d). As expected, the BIC resonance is much sharper than those of the plasmonic array discussed above, yielding much smaller decay rates, as we show below (see also the SM \cite{supplemental_material} for further details).

 \begin{figure}[b]
    \centering
    \includegraphics[width=1\linewidth]{Fig4.pdf}
    \caption{\textit{Resonant quantum light generation from crystal-polariton nonlinearities.} (a) Schematic of the proposed setup. (b) Two-photon emission rate density $\Gamma_{\vec{k}_\parallel} $ normalized by the incoming flux density $\Phi_\mathrm{in}^2$ ($\Gamma_{\vec{k}_\parallel} \propto \Phi_\mathrm{in}^2$ in the perturbative regime) for $k_\mathrm{L} = 6.4\times 10^{-3} \pi/a$ (vertical dotted line) and $V_d/V_B = 1.6\times 10^{-5}$ as a function of laser frequency $\omega_L$ and in-plane momentum $\vec{k}_\parallel = k_\parallel  \vec{e}_y $ of one of the emitted photons (in-plane momentum of the other emitted photon is fixed to $2k_\mathrm{L}-k_\parallel$ such that the plot is symmetric around $k_\mathrm{L}$). Enhanced two-photon emission occurs when the laser frequency is in resonance with the upper or lower crystal polaritons at the laser momentum $\omega_\mathrm{UP/LP}$ and the resonance condition $\omega(\vec{k}_\parallel) + \omega(\overline{\vec{k}}_\parallel) = 2 \omega_L$, shown by white dashed lines, is satisfied.}
    \label{fig:Nonlinear}
\end{figure}

We find that the BIC resonance in the reciprocal-space spectral density can be modeled by a single lossy mode for each value of the in-plane wavevector; see \autoref{fig:BIC}(c). We further obtain the decay rate and coupling strength of this mode to an emitter with a dipole moment of $10\,$D and a frequency of $\omega_E = 1.749\,$eV by a single-mode fit and show them as functions of in-plane momentum in \autoref{fig:BIC}(d). Although the decay rate vanishes when approaching the $\Gamma$ point at $\vec{k}_\parallel = \vec{0}$, leading to a divergent quality factor, we find that the coupling strength remains finite and essentially constant over the resolved wave-vector range. Clearly, we find $g>\kappa$, such that the system operates in the strong coupling regime. To confirm this, we show the electric field at the position of the emitter under plane-wave driving with different in-plane momenta and frequencies in \autoref{fig:BIC}(b); see SM for details \cite{supplemental_material}. A clear anticrossing can be found that indicates the strong-coupling regime. The dispersion of these polariton modes can be obtained by diagonalizing the single excitation manifold of the few-mode Hamiltonian [white dashed line in \autoref{fig:BIC}(b)]. This offers a highly tunable strong-coupling platform, as many different crystal polaritons can be realized depending on the lattice geometry and the position of the emitter.

\paragraph{Resonant quantum nonlinear optics with crystal polaritons.}

The high $Q$ resonances of dielectric metasurfaces have already been shown to boost nonlinear light conversion in a series of experimental works relying on off-resonant light-matter interactions \cite{liu2019high,koshelev2019nonlinear,yang2015nonlinear,grinblat2021nonlinear}. Because the matter
component of crystal polaritons is composed of two-level excitations, these hybrid modes inherit
a strong quantum nonlinearity. As a first application of crystal polaritons, we therefore show how driving them with a laser can lead to resonant nonlinear light generation with an increase in the nonlinear light conversion by many orders of magnitude compared to conventional platforms based on bulk nonlinearities \cite{liu2019high,koshelev2019nonlinear,yang2015nonlinear,grinblat2021nonlinear}. 

In the low-excitation regime, we can approximate the two-level emitters by harmonic oscillators with annihilation operators $\hat{b}_{\vec{k}_\parallel}$. As a result, $\vec{k}_\parallel$ is conserved on the single particle level. We introduce the nonlinearity of the two-level emitters via the resummation of the Holstein--Primakoff expansion \cite{holstein1940field,vogl2020resummation} of $\hat{\sigma}_{\vec{k}_\parallel}$ in terms of $\hat{b}_{\vec{k}_\parallel}$, resulting in a polariton-polariton scattering term in the Hamiltonian \cite{supplemental_material}. A strong nonlinear response with just one excitation in the system arises when this term induces a frequency shift of the two-polariton state that is larger than its linewidth \cite{chang2014quantum,delteil2019towards,munoz2019emergence}. For our system, we find that this is satisfied for polariton densities as low as one polariton per $10.7\, \mu \mathrm{m}^{2}$ (=$67$ unit cells) \cite{supplemental_material}, and should also be compared to the free-space wavelength of $0.71\mu$m of the laser driving. This demonstrates the potential of our setup for quantum optical applications, which we will explicitly demonstrate using the example of the efficient generation of directionally emitted entangled photon pairs [see Fig.~\ref{fig:Nonlinear}(a)].

To this end, we introduce laser driving with fixed frequency $\omega_L$ and in-plane momentum $\vec{k}_L$, see \autoref{fig:Nonlinear}(a), accounting for light scattering at the metasurface self-consistently following Refs.~\cite{feist_macroscopic_2020,lindel_macroscopic_2021}, as outlined in the SM \cite{supplemental_material}. In the low-excitation regime and in the steady state the laser driving coherently displaces the modes and the emitters with momentum $\vec{k}_L$ by $\mathcal{a}_{\vec{k}_L,i} $ and $\mathcal{b}_{\vec{k}_L} $, respectively, while all emitter and photon states with $\vec{k}_\parallel \neq \vec{k}_L$ remain in the ground state. Assuming that the nonlinearity is weak, we find that the lowest-order correction affecting modes with $\vec{k}_\parallel \neq \vec{k}_\mathrm{L}$ is given by a squeezing and a beam-splitter contribution \cite{supplemental_material}. The beam-splitter contribution leads to a negligible shift of the mode and emitter frequencies, while the squeezing Hamiltonian leads to the generation of entangled excitations and reads $ \hat{H}_{\mathrm{sq},\vec{k}_\parallel} = \sum_{i} g_{\vec{k}_\parallel,i} ^\ast\mathcal{b}_{\vec{k}_\mathrm{L}}^2 \hat{a}_{\vec{k}_\parallel,i}^\dagger \hat{b}^\dagger_{\kparb}  + E_\mathrm{loc}\mathcal{b}_{\vec{k}_\mathrm{L}} \hat{b}_{\vec{k}_\parallel}^\dagger \hat{b}_{\kparb}^\dagger + (\vec{k}_\parallel \leftrightarrow \kparb) + \mathrm{h.c.}$, where $ (\vec{k}_\parallel \leftrightarrow \kparb) $ denotes adding the preceding terms subject to the replacement $\vec{k}_\parallel \leftrightarrow \kparb $ with $\kparb \equiv 2\vec{k}_\mathrm{L} - \vec{k}_\parallel$.
Allowing for up to two excitations in the system, we can numerically identify the steady-state two-photon emission rate at a fixed wavevector $\vec{k}_\parallel$ defined by $
    \gamma_{\vec{k}_\parallel} = \sum_{i,j} ( \kappa_{\vec{k}_\parallel, i} +\kappa_{\overline{\vec{k}}_\parallel, j} )\braket{\hat{a}_{\vec{k}_\parallel, i}^\dagger \hat{a}_{\overline{\vec{k}}_\parallel, j}^\dagger \hat{a}_{\vec{k}_\parallel, i} \hat{a}_{\overline{\vec{k}}_\parallel, j}}_\mathrm{ss}.$ 
Here, $\braket{\dots}_\mathrm{ss}$ is the steady-state expectation value. The two-photon emission rate density per unit area into a certain range of wave-vectors $\vec{k}_\parallel \in V_d$ is defined by $\Gamma_{V_d} = 1/(4\pi^2) \int_{V_d} \dif^2 k_\parallel   \gamma_{\vec{k}_\parallel}$ and can be approximated by $ \Gamma_{\vec{k}_\parallel} = V_d/(4\pi^2)  \gamma_{\vec{k}_\parallel}$ for small $V_d$. %, where $V_B = (2\pi/a)^2 $ is the Brillouin zone volume. 
Note that since the squeezing Hamiltonian $\hat{H}_{\mathrm{sq},\vec{k}_\parallel} $ always generates pairs of excitations at $\vec{k}_\parallel$ and $\overline{\vec{k}}_\parallel$, the emitted photons are strongly entangled and we only have to integrate over a single wave-vector $\vec{k}_\parallel$, while the wave-vector of the other emitted photon is fixed by momentum conservation to $2\vec{k}_\mathrm{L}-\vec{k}_\parallel$. 

In \autoref{fig:Nonlinear}(b), we plot $\Gamma_{\vec{k}_\parallel}$ for $k_L = 6.4\times   10^{-3}  \pi/a $ as a function of the laser driving frequency $\omega_L$ and the in-plane momentum of one of the emitted photons $\vec{k}_\parallel$. The maxima in the two-photon emission rates can be understood as a result of two resonant conditions: (i) the laser driving has to be resonant with one of the polaritonic modes $\omega_\mathrm{UP}$ and $\omega_\mathrm{LP}$ at $\vec{k}_L$ (the maxima in \autoref{fig:Nonlinear}(b) are clearly centered around $\omega_\mathrm{UP}$ and $\omega_\mathrm{LP}$); (ii) and momentum (phase-matching) and energy must be conserved, i.e. there must be polaritonic modes with frequencies $\omega(\vec{k}_\parallel)$ and $ \omega(\overline{\vec{k}}_\parallel)$ satisfying $\omega(\vec{k}_\parallel) + \omega(\overline{\vec{k}}_\parallel) = 2 \omega_L$. These conditions are plotted by the white dashed lines in \autoref{fig:Nonlinear}(b). As a result of both resonant conditions, we see maxima in the two-photon emission rate when the dashed lines cross $\omega_\mathrm{LP}$ or $\omega_\mathrm{UP}$ in \autoref{fig:Nonlinear}(b). We find that the normalized two-photon emission rate $\Gamma_{\vec{k}_\parallel}/\Phi_\mathrm{in}^2$ emitted into a narrow fraction $V_d/V_\mathrm{B}= 1.6\times 10^{-5}$ of the Brillouin zone $V_B = (2\pi/a)^2 $ can reach values up to $10^{-2}\,\mathrm{cm}^2\mathrm{fs}$. This means that for an incoming laser power density of, e.g., just $ P = 10\,\mathrm{\mu W}/\mathrm{cm}^2$ the conversion efficiency into that narrow range of wave-vectors is $\Gamma_{\vec{k}_\parallel}/\Phi_\mathrm{in} = 3.6\times 10^{-4}$ and for $P> 28 \mathrm{mW}/\mathrm{cm}^2$ the system would already be far in the non-perturbative regime (perturbative conversion efficiency $>1$). Nonlinear metasurfaces with dielectric nanoparticles based on bulk nonlinearities \cite{liu2019high,koshelev2019nonlinear,yang2015nonlinear,grinblat2021nonlinear} have reached conversion efficiencies of $10^{-6}$, however, requiring laser powers of $60\,\mathrm{MW/cm}^2$, which makes our platform more efficient by $14$ orders of magnitude.

\paragraph{Conclusion and outlook.}

In summary, we have shown that metasurfaces can be promoted from classical wave-shaping elements to a tuneable cavity-QED platform for periodic quantum matter. Treating photonic Bloch resonances and emitter-array spin waves on equal footing, we have shown that they can hybridize into crystal polaritons with a nonlinear optical response that can be used for efficient quantum light generation. The complete setup can be seen as a polaritonic quantum metasurface with many different design possibilities. Here we have focused on wavelength-sized arrays, but sub-wavelength arrays are also expected to offer great potential for quantum optics \cite{shahmoon2017cooperative,perczel2017photonic,rui2020subradiant,bettles2020quantum,moreno2021quantum,zhang2022photon,pedersen2023quantum,pedersen2024green,asenjo2017exponential,bluvstein2024logical,reitz2022cooperative}. Moreover, chiral electromagnetic modes, whose design in typical cavity QED platforms is challenging and which are of particular interest in chiral sensing and nonreciprocal quantum devices, as well as topologically nontrivial states of finite-sized metasurfaces, could be harnessed in the future by the proposed polaritonic quantum metasurface.

\acknowledgments{
     We thank Jaime Abad-Arredondo and Alberto Miguel-Torcal for fruitful discussions. We acknowledge financial support by the Spanish Ministerio de Ciencia y Universidades---Agencia Estatal de Investigación through grants  PID2021-125894NB-I00, EUR2023-143478, and CEX2023-001316-M through the María de Maeztu program for Units of Excellence in R\&D, and from the European Union's Horizon Europe research and innovation programme under Grant Agreements No. 101070700 (MIRAQLS) and No. 101098813 (SCOLED). The work was further supported by the ETH Quantum Center Research Fellowship and the Dr Alfred and Flora Spälti Fonds.}

\newpage\clearpage
% === SUPPLEMENTAL MATERIAL ===

%\begin{bibunit}

\setcounter{secnumdepth}{2}
\setcounter{page}{1}
\setcounter{equation}{0}
\renewcommand{\thesection}{S\Roman{section}}
\renewcommand{\theequation}{S\arabic{equation}}
\renewcommand{\thefigure}{S\arabic{figure}}
\renewcommand{\thetable}{S\arabic{table}}

\onecolumngrid

\begin{center}
  \textbf{\large Supplemental Material}\\[1em]
  \normalsize (for the article: “Close encounters between periodic light and periodic arrays of quantum emitters”)\\[2em]
\end{center}

\section{Light-matter interaction in 2D periodic photonic structures}

A formal quantization of the electromagnetic field in arbitrary structures, including dispersive and lossy media, is achieved by the macroscopic QED formalism \cite{scheel_macroscopic_2008,feist_macroscopic_2020}. This framework describes the electromagnetic field through a set of bosonic vector operators $\left\lbrace \hat{\vec{f}}_{p}(\vec{r},\omega)\right\rbrace$ projected on the dyadic Green tensor $\tens{G}\left(\vec{r},\vec{r'},\omega\right)$. The electric field operator reads
\begin{align}
    \hat{\vec{E}}\left(\vec{r}\right) = \sum_{p=e,m}\int_0^{\infty}d\omega\int d^3r^{\prime}  
  \left[\tens{G}_{p}\left(\vec{r},\vec{r'},\omega\right)\cdot \hat{\vec{f}}_{p}\left(\vec{r'},\omega\right) + \textup{h}.\textup{c}. \right],
\end{align}
where the amplitude coefficients $\tens{G}_{p}\left(\vec{r},\vec{r'},\omega\right)$ are related to the dyadic Green tensor by 
\begin{align}%\label{eqapp:coefficients_mqed}
        \tens{G}_{e}(\vec{r},\vec{r'},\omega)&=i\frac{\omega ^2}{c^2}\sqrt{\frac{\hbar}{\pi \epsilon _0} \textup{Im} \hspace{0.04cm} \epsilon (\vec{r'},\omega)} \hspace{0.05cm} \tens{G}(\vec{r},\vec{r'},\omega), \label{eq:Ge}
        \\ \label{eq:Gm}
        \tens{G}_{m}(\vec{r},\vec{r'},\omega)&=i\frac{\omega}{c}\sqrt{\frac{\hbar}{\pi \epsilon _0} \frac{\textup{Im} \hspace{0.04cm} \mu(\vec{r'},\omega)}{{\lvert \mu(\vec{r'},\omega)\rvert}^2} }   \times \left[\overrightarrow\nabla_{r'}\times \tens{G}(\vec{r},\vec{r'},\omega) \right]^{T}, 
\end{align}
where $\epsilon(\vec{r},\omega)$ is the electric permittivity and $\mu(\vec{r},\omega)$ is the magnetic permeability. As we consider a 2D periodic photonic structure, these two quantities fulfill the periodicity conditions $\epsilon(\vec{r}+\vec{R},\omega)=\epsilon(\vec{r},\omega)$ and $\mu(\vec{r}+\vec{R},\omega)=\mu(\vec{r},\omega)$, with $\vec{R}$ being an arbitrary lattice vector. These properties translate into an analogous periodicity condition for the Green tensor:
\begin{equation}\label{eqapp:periodic_condition}
    \tens{G}\left(\vec{r} + \vec{R},\vec{r'} + \vec{R},\omega\right) = \tens{G}\left(\vec{r},\vec{r'},\omega\right).
\end{equation}

Within this quantization scheme, the total Hamiltonian of our periodic photoninc structure coupled to the emitter array is written as
\begin{equation}
    \hat{H} = \hat{H}_{e} + \sum_{p=e,m}\int_0^{\infty}d\omega\int d^3r\hspace{0.05cm}\hbar\omega\hspace{0.03cm}\hat{\vec{f}}_{p}^{\dagger}\left(\vec{r},\omega\right)\cdot\hat{\vec{f}}_{p}\left(\vec{r},\omega\right)
    - \sum_{i}\sum_{h=1}^{N_E}\hat{\vec{d}}_i^{(h)}\cdot\hat{\vec{E}}(\vec{r}_i^{(h)}),
\end{equation}
where $\hat{H}_{e}$ is the Hamiltonian of the $2$D-periodic array of emitters and $\hat{\vec{d}}_i^{(h)}=\hat{d}_i^{(h)}\vec{n}^{(h)}$ is the dipole operator of the multilevel emitter $h$ in the unit cell $i$, located at $\vec{r}_{i}^{(h)}=\vec{r}_{0}^{(h)}+\vec{R}_{i}$, $\vec{r}_{0}^{(h)}$ being the position of the emitter in the reference unit cell. Note that we consider a single orientation $\vec{n}^{(h)}$ per emitter and did not perform the two-level approximation here.

The periodic symmetry of the system implies that Bloch's theorem holds and, as shown in the following, that the full problem can be reduced to a problem restricted to the reciprocal unit cell.

To do so, we define the spin-wave basis for the emitter degrees of freedom
\begin{subequations}\label{eqapp:ft_discrete}
    \begin{align}
        \hat{d}^{(h)}_{\vec{k}_{\parallel}} & = \sum_{i} \hat{d}_{i}^{(h)}e^{-i\vec{k}_{\parallel}\cdot\vec{r}_{i}^{(h)}},
        \\
        \hat{d}_{i}^{(h)} & = \frac{1}{V_B}\int_{V_B}d^2k_{\parallel}\hspace{0.04cm}\hat{d}^{(h)}_{\vec{k}_{\parallel}}e^{i\mathbf{k}_{\parallel}\cdot\vec{r}_{i}^{(h)}}.
    \end{align}
\end{subequations}
In addition, defining a reciprocal-space analogue of emitter-centered modes \cite{buhmann_casimir-polder_2008,feist_macroscopic_2020}
\begin{equation}
    \hat{B}_{\vec{k}_{\parallel}}^{(h)}(\omega)=\frac{1}{\hbar}\sum_ie^{-i\vec{k}_{\parallel}\cdot\vec{r}_{i}^{(h)}}
    \sum_{p}\int d^3r\hspace{0.05cm}\vec{n}^{(h)}\cdot\tens{G}_p(\vec{r}_i^{(h)},\vec{r},\omega)\cdot\hat{\vec{f}}(\vec{r},\omega),
\end{equation}
we find that the interaction Hamiltonian reads
\begin{align}
      \hat{H}_I = -\frac{\hbar}{V_B}\int_0^{\infty}d\omega\int_{V_B}d^2k_{\parallel} \sum_h  \hat{d}^{(h)\dagger}_{\vec{k}_{\parallel}}\hat{B}_{\vec{k}_{\parallel}}^{(h)}(\omega)+ \mathrm{h.c.}
\end{align}

The commutation relations 
\begin{align}
    \left[\hat{B}_{\vec{k}_{\parallel}}^{(h)}(\omega), \hat{B}_{\vec{k}_{\parallel}^{\prime}}^{(h^{\prime})\dagger}(\omega^{\prime})\right] 
    =V_B\mathcal{J}_{h,h^{\prime}}(\mathbf{k}_{\parallel},\omega)\delta(\omega-\omega^{\prime})\delta(\mathbf{k}_{\parallel}-\mathbf{k}_{\parallel}^{\prime}),
\end{align} 
determine the reciprocal-space spectral density that fully describes the light-matter coupling (see Eq.~\eqref{eqapp:HamiltonianRec}):
\begin{align} \label{eqapp:spectraldensity}
\mathcal{J}_{h,h^\prime}(\vec{k}_\parallel, \omega) = \frac{ \mu_0 \omega^2 }{\pi \hbar}  \me^{-\mi \vec{k}_\parallel (r_0^{(h)}-r_0^{(h^\prime)})}
\vec{n}^{(h)} \cdot  \left[\sum_{i}\me^{\mi \vec{k}_\parallel \cdot \vec{R}_i}\mathrm{Im}\hspace{0.03cm}\tens{G}(\vec{r}_0^{(h)},\vec{r}_0^{(h^\prime)}+\vec{R}_i,\omega)\right]\cdot \vec{n}^{(h^\prime)}.
\end{align}
In the derivation, we have used the important relation \cite{scheel_macroscopic_2008,Buhmann2012a}
\begin{align}\label{eqapp:GT_formula}
    \sum_{p}\int d^3s\hspace{0.05cm}\tens{G}_{p}(\vec{r},\vec{s},\omega)\cdot\tens{G}_{p}^{\dagger}(\vec{r}^{\prime},\vec{s},\omega)
    =\frac{\hbar\mu_0\omega^2}{\pi}\mathrm{Im}\hspace{0.03cm}\tens{G}(\vec{r},\vec{r}^{\prime},\omega),
\end{align}
as well as the periodicity condition in Eq.~\eqref{eqapp:periodic_condition} and the completeness condition for plane waves $\sum_{i}\me^{\mi(\vec{k}_{\parallel}-\vec{k}_{\parallel}^{\prime})\cdot\vec{R}_i}=V_{B}\delta(\vec{k}_{\parallel}-\vec{k}_{\parallel}^{\prime})$. By comparing the reciprocal-space spectral density in Eq.~\eqref{eqapp:spectraldensity} to its real-space analogue \cite{medina_few-mode_2020}, we find that the imaginary part of the dyadic Green tensor found in the latter is replaced by the lattice sum of the imaginary part of the dyadic Green tensor in the former. Furthermore, note that the diagonal elements of $\boldsymbol{\mathcal{J}}(\mathbf{k}_{\parallel},\omega)$ are equal to the conventional single-transition spectral density $\boldsymbol{J}(\mathbf{k}_{\parallel},\omega)$ given in  the main text up to the square of the transition dipole moment, $d_{h}^{2}\mathcal{J}_{h,h} = J_{h,h}$. The same argument is true for the cross terms, $d_{h}d_{h^{\prime}}\mathcal{J}_{h,h^{\prime}} = J_{h,h^{\prime}}$.

Further insight into Eq~\eqref{eqapp:spectraldensity} can be provided by recalling the definition of the Bloch-periodic Green tensor, $ \boldsymbol{\mathcal{G}} (\vec{k}_\parallel, \vec{r},\vec{r}^\prime ,\omega) =  \sum_{i} \me^{\mi \vec{k}_\parallel \cdot \vec{R}_i}  \tens{G} ( \vec{r},\vec{r}^\prime+\vec{R}_i ,\omega)$, which is the solution of Maxwell's equations with Bloch-periodic boundary conditions. Equation~\eqref{eqapp:periodic_condition} implies that Bloch's theorem is fulfilled such that $\boldsymbol{\mathcal{G}} (\vec{k}_\parallel, \vec{r}+\vec{R}_i,\vec{r}^\prime ,\omega)=\me^{\mi \vec{k}_\parallel \cdot \vec{R}_i}\boldsymbol{\mathcal{G}} (\vec{k}_\parallel, \vec{r},\vec{r}^\prime ,\omega)$. We can then use the Bloch-periodic Green tensor and Eq.~\eqref{eqapp:periodic_condition} to find that the anti-Hermitian part of the Bloch-periodic Green tensor reads
\begin{align}
   \boldsymbol{\mathcal{G}}_\mathrm{AH}(\vec{k}_\parallel, \vec{r}_0^{(h)},\vec{r}_0^{(h^\prime)} ,\omega)&
   \equiv \frac{1}{2\mi}\left[\boldsymbol{\mathcal{G}}(\mathbf{k}_{\parallel},\mathbf{r}_0^{(h)},\mathbf{r}_0^{(h')},\omega)-\boldsymbol{\mathcal{G}}^{\dagger}(\mathbf{k}_{\parallel},\mathbf{r}_0^{(h')},\mathbf{r}_0^{(h)},\omega)\right] \\
  & =  \sum_{i}\me^{\mi \vec{k}_\parallel \cdot \vec{R}_i}\mathrm{Im}\hspace{0.03cm}\tens{G}(\vec{r}_0^{(h)},\vec{r}_0^{(h^\prime)}+\vec{R}_i,\omega),
\end{align} 
such that Eq.~\eqref{eqapp:spectraldensity} reduces to Eq.~(1) of the main text.

The next step is to perform an orthogonalization transformation $\hat{B}_{\vec{k}_{\parallel}}^{(h)}(\omega)=\sum_{h^{\prime}}W_{h,h^{\prime}}(\vec{k}_{\parallel},\omega)\hat{a}_{\vec{k}_{\parallel}}^{(h^{\prime})}(\omega)$ such that $\left[\hat{a}_{\vec{k}_{\parallel}}^{(h)}(\omega),\hat{a}_{\vec{k}_{\parallel}^{\prime}}^{(h^{\prime})\dagger}(\omega^{\prime})\right]=V_{B}\delta_{hh^{\prime}}\delta(\vec{k}_{\parallel}-\vec{k}_{\parallel}^{\prime})\delta(\omega-\omega^{\prime})$, where $\tens{W}\cdot\tens{W}^{\dagger}=\boldsymbol{\mathcal{J}}$ \cite{sanchez2022few}. The interaction Hamiltonian then given by
\begin{align}
    H_I = -\frac{\hbar}{V_{B}}\int_{V_B} d^2k_{\parallel}\int_0^{\infty}d\omega\sum_{h,h^{\prime}}
    \left[\hat{d}^{(h)\dagger}_{\vec{k}_{\parallel}}W_{h,h^{\prime}}(\vec{k}_{\parallel},\omega)\hat{a}_{\vec{k}_{\parallel}}^{(h^{\prime})}(\omega) +\mathrm{h.c.}\right].
\end{align}

The total Hamiltonian can be finally written in terms of the momentum resolved Hamiltonian $\hat{H}_{\vec{k}_{\parallel}}$:
\begin{equation}
    \hat{H}=\frac{1}{V_B}\int_{V_B}d^2k_{\parallel}\hat{H}_{\vec{k}_{\parallel}},
\end{equation}
with:
\begin{align} \label{eqapp:HamiltonianRec}
    \hat{H}_{\vec{k}_{\parallel}} = \hat{H}_{e,\vec{k}_{\parallel}} + \int_{0}^{\infty}d\omega\sum_{h}\hbar\omega\hspace{0.05cm}\hat{a}_{\vec{k}_{\parallel}}^{(h)\dagger}(\omega)\hat{a}_{\vec{k}_{\parallel}}^{(h)}(\omega)
    -\int_0^{\infty}d\omega\sum_{h,h^{\prime}}\left[\hat{d}^{(h)\dagger}_{\vec{k}_{\parallel}}\hbar W_{h,h^{\prime}}(\vec{k}_{\parallel},\omega)\hat{a}_{\vec{k}_{\parallel}}^{(h^{\prime})}(\omega)+\mathrm{h.c.}\right].
\end{align}

To find the reciprocal-space spectral density for the configuration considered in the main text, we used the $T$-matrix-based scattering code \textit{treams} \cite{beutel2024treams}.

From now on, we set $\hbar = 1$.

\section{Few-mode model dynamics}

As described in the main text, one can follow Refs.~\cite{tamascelli2018nonperturbative,lentrodt2020ab,medina_few-mode_2020,sanchez2022few,sanchez2024mixed,lednev2025spatially} to construct a few-mode model in which the continuum of field modes seen by the emitters is equivalently described by a set of few coupled lossy field modes given by creation and annihilation operators $\hat{a}_{\vec{k}_\parallel,i}^{(\dagger)}$. The density matrix $\hat{\rho}$ of the emitters and the modes then evolves as
$
\dot{\hat{\rho}} =(1/V_\mathrm{B}) \int_{V_\mathrm{B}}\dif^2 k_\parallel(-\mi) [\hat{H}_{\vec{k}_\parallel} , \hat{\rho}  ] + \sum_{i} \kappa_{\vec{k}_\parallel,i} L_{\hat{a}_{\vec{k}_\parallel,i}}[\hat{\rho}]$,
with the Lindblad dissipator $L_{\hat{O}}[\hat{\rho}] = \hat{O} \hat{\rho} \hat{O}^\dagger - 
\frac{1}{2}\{\hat{O}^\dagger \hat{O}, \hat{\rho} \}$, the Brillouin zone volume of the lattice $V_\mathrm{B}$, and the few-mode Hamiltonian $\hat{H}_{\vec{k}_\parallel}$ given in Eq.~(2) of the main text that we repeat here for clarity:
\begin{align}   \label{eq:few_mode_hamiltonian}
 \hat{H}_{\vec{k}_\parallel} =  \hat{H}_{e,\vec{k}_\parallel} + \sum_{i,j} \omega_{\vec{k}_\parallel,ij}  \hat{a}_{\vec{k}_\parallel,i}^\dagger \hat{a}_{\vec{k}_\parallel,j}   +   \sum_{i,h} ( g_{ \vec{k}_\parallel,ih} \hat{d}_{\vec{k}_\parallel}^{(h)\dagger}    \hat{a}_{\vec{k}_\parallel,i} + \mathrm{h.c.}) .
\end{align}

\section{Laser driving} \label{sec:Driving}

In the framework of macroscopic quantum electrodynamics, coherent laser driving in the present of absorptive and lossy environments can be included straightforwardly \cite{feist_macroscopic_2020,lindel_macroscopic_2021}. Formally, the emitters and field modes are driven by the laser. One can analytically remove the driving of the field modes, by replacing the driving field of the emitters by the structure-enhanced field at the position of the $h$th emitter in the unit cell $\vec{E}_h$, i.e. by the field obtained by propagating the incoming laser field to the location of the emitter $\vec{r}_0^{(h)}$ in the presence of the electromagnetic environment (here the metasurface). 

In the manuscript, we consider single-mode plane wave driving with frequency $\omega_\mathrm{L}$ and in-plane wave-vector $\vec{k}_\mathrm{L}$ and find $\vec{E}_h = \widetilde{\vec{E}}_h \me^{-\mi \vec{k}_\mathrm{L} \cdot \vec{r}_0^{(h)}}$ numerically again using the $T$-matrix-based scattering code \textit{treams} \cite{beutel2024treams}. The corresponding driving term in the Hamiltonian then reads 
\begin{align}
    \hat{H}_\mathrm{L} = \sum_{h}\Omega_h (\hat{\sigma}^{(h)\dagger}_{\vec{k}_\mathrm{L}}+ \hat{\sigma}^{(h)}_{\vec{k}_\mathrm{L}}   ),
\end{align}
where the Rabi frequency of the laser drive is given by $\Omega_h = \vec{d}_h \cdot \widetilde{\vec{E}}_h$.

\section{Holstein-Primakoff transformation and squeezing Hamiltonian}

In general, the equations of motion at fixed in-plane momentum of the few-mode model,
\begin{align} \label{eq:eom}
\dot{\hat{\rho}}_{\vec{k}_\parallel} = -\mi [\hat{H}_{\vec{k}_\parallel} , \hat{\rho}_{\vec{k}_\parallel}  ] + \sum_{i} \kappa_{\vec{k}_\parallel,i} L_{\hat{a}_{\vec{k}_\parallel,i}}[\hat{\rho}_{\vec{k}_\parallel}],
\end{align}
with $L_{\hat{O}}[\hat{\rho}] = \hat{O} \hat{\rho} \hat{O}^\dagger - 
\frac{1}{2}\{\hat{O}^\dagger \hat{O}, \hat{\rho} \}$,
are not closed, since spin-wave excitations with different in-plane momenta do not commute, that is, $[\sigma_{\vec{k}_\parallel}^{(h)},\sigma_{\vec{k}_\parallel^\prime}^{(h) \dagger}] \neq 0$ for $\vec{k}_\parallel \neq \vec{k}_\parallel^\prime$. In this section, we simplify the equations of motion of the few-mode model by systematically performing a low-excitation approximation via a Holstein-Primakoff transformation \cite{holstein1940field}
\begin{align} \label{eqapp:holsteinprimakoff}
    \hat{\sigma}_{i} = \sqrt{1-\hat{b}_{i}^{\dagger}\hat{b}_{i}}\hat{b}_{i}.
\end{align}
$\hat{b}_{i}$ are bosonic annihilation operators and note that, for simplicity, here we only consider the case of a single emitter per unit cell with $ \hat{\sigma}_{\vec{k}_\parallel}^{(1)} =  \hat{\sigma}_{\vec{k}_\parallel} $.

\subsection{Linear approximation}%teady-state, and field seen by the emitter}

In the low-excitation regime, we have $\braket{\hat{b}_{i}^{\dagger}\hat{b}_{i} } \ll 1$ such that we can expand Eq.~\eqref{eqapp:holsteinprimakoff} to lowest order, finding $ \hat{\sigma}_{i} \approx \hat{b}_{i}$. In reciprocal space, this reads $ \hat{\sigma}_{\vec{k}_\parallel} \approx \hat{b}_{\vec{k}_\parallel}$ with
\begin{align}
    \hat{b}_{\vec{k}_\parallel} = \sum_{i} \hat{b}_{i} \me^{-\mi \vec{k}_\parallel \cdot \vec{r}_{i}},
\end{align}
where
\begin{align}
    [\hat{b}_{\vec{k}_\parallel},\hat{b}_{\vec{k}_{\parallel^{\prime}}}^{\dagger}] = V_B \delta(\vec{k}_\parallel-\vec{k}_\parallel^\prime).
\end{align}
As a result, we find the linearized few-mode Hamiltonian with $N_m$ photonic modes  
\begin{align} \label{eqapp:H_linear}
\hat{H}_{\vec{k}_\parallel, \mathrm{lin}} = \omega_\mathrm{E}  \hat{b}_{\vec{k}_\parallel}^\dagger \hat{b}_{\vec{k}_\parallel} +  \sum_{i,j=1}^{N_m} \omega_{\vec{k}_\parallel,ij}\hat{a}_{\vec{k}_\parallel,i}^\dagger \hat{a}_{\vec{k}_\parallel,j}  +   \sum_{i=1}^{N_m} ( g_{ \vec{k}_\parallel,i} \hat{b}_{\vec{k}_\parallel}^{\dagger}    \hat{a}_{\vec{k}_\parallel,i} + \mathrm{h.c.}) ,
\end{align}
which conserves the in-plane momentum on the single particle level, such that we can diagonalize it for each $\vec{k}_\parallel$ individually. Diagonalization results in polaritonic operators $\hat{\vec{P}}_{\vec{k}_\parallel} = (\hat{P}_{\vec{k}_\parallel, 1}, \dots, \hat{P}_{\vec{k}_\parallel, N_m+1})^{\mathrm{T}} $ that are related to $\hat{\vec{A}}_{\vec{k}_\parallel} = (\hat{a}_{\vec{k}_\parallel,1}, \dots, \hat{a}_{\vec{k}_\parallel,N_m},\hat{b}_{\vec{k}_\parallel} )^{\mathrm{T}} $ via a linear transformation $\hat{\vec{P}}_{\vec{k}_\parallel}  = \tens{V}_{\vec{k}_\parallel} \cdot \hat{\vec{A}}_{\vec{k}_\parallel}$ and the Hamiltonian in the polaritonic basis reads
\begin{align}
    \hat{H}_{\vec{k}_\parallel, \mathrm{lin}} =  \sum_{i} \omega_{\vec{k}_\parallel,i}^{(\mathrm{P})}\hat{P}_{\vec{k}_\parallel,i}^\dagger \hat{P}_{\vec{k}_\parallel,i}.
\end{align}
The polaritonic frequencies $\omega_{\vec{k}_\parallel,i}^{(\mathrm{P})}$ for the bound state in the continuum strongly coupled to an array of quantum emitters considered in the main text are shown in Fig.~3(b) of the main text by the white dashed lines.

\subsection{Polariton-polariton scattering}

To account for the nonlinearity of the two-level emitters, we consider the resummation \cite{vogl2020resummation} of the Holstein-Primakoff expansion \eqref{eqapp:holsteinprimakoff} which is exact for single spins considered here, by setting
\begin{align}
\hat{\sigma}_i \approx \left(1- \hat{b}_i^\dagger \hat{b}_i  \right) \hat{b}_i.
\end{align}
In reciprocal space, we find 
\begin{align} \label{eqapp:HolsteinPrimakoff_kspace}
     \hat{\sigma}_{\vec{k}_\parallel} = \hat{b}_{\vec{k}_\parallel} - \frac{1}{V_B^2} \int_{V_B} \dif^2 k_\parallel^\prime \int_{V_B}\dif^2 k_\parallel^{\prime\prime}   \,\hat{b}^\dagger_{\vec{k}_\parallel^\prime}\hat{b}_{\vec{k}_\parallel^{\prime\prime}} \hat{b}_{\vec{k}_\parallel + \vec{k}^\prime_\parallel - \vec{k}_\parallel^{\prime\prime}}.
\end{align}
As a result, there is an additional nonlinear contribution to the linear Hamiltonian in Eq.~\eqref{eqapp:H_linear}
\begin{align} \label{eqapp:HNL}
    H_\mathrm{NL} = - \frac{1}{V_B^3} \int_{V_B} \dif^2 k_\parallel \int_{V_B} \dif^2 k_\parallel^\prime \int_{V_B}\dif^2 k_\parallel^{\prime\prime}   \, \sum_{i=1}^{N_m} g_{\vec{k}_\parallel,i} \hat{b}^\dagger_{\vec{k}_\parallel^\prime}\hat{b}^\dagger_{\vec{k}_\parallel^{\prime\prime}} {\hat{b}_{\vec{k}_\parallel^{\prime}  + \vec{k}^{\prime\prime}_\parallel - \vec{k}_\parallel}} a_{\vec{k}_\parallel,i}  + \mathrm{h.c.},
\end{align}
which can also be expressed in terms of the polaritonic operators
\begin{multline}
       H_\mathrm{NL} = - \frac{1}{V_B^3} \int_{V_B} \dif^2 k_\parallel \int_{V_B} \dif^2 k_\parallel^\prime \int_{V_B}\dif^2 k_\parallel^{\prime\prime}   \, \sum_{i=1}^{N_m} \sum_{j,l,m,n = 1}^{N_m+1}\\
       \times g_{\vec{k}_\parallel,i}  {V_{\vec{k}^{\prime}_\parallel,j,N_m+1}V_{\vec{k}^{\prime\prime}_\parallel,l,N_m+1}V^\ast_{\vec{k}^{\prime}_\parallel+\vec{k}^{\prime\prime}_\parallel-\vec{k}_\parallel,m,N_m+1} V^\ast_{\vec{k}_\parallel,n,i} }\hat{P}^\dagger_{\vec{k}_\parallel^\prime,j}\hat{P}^\dagger_{\vec{k}_\parallel^{\prime\prime},l} {\hat{P}_{\vec{k}_\parallel^{\prime} + \vec{k}^{\prime\prime}_\parallel - \vec{k}_\parallel ,m}} \hat{ P}_{\vec{k}_\parallel,n}  + \mathrm{h.c.}
\end{multline}
For one fixed in-plane momentum, all polaritonic states with a different in-plane momentum can be seen as a reservoir to which the singled-out momentum polaritons couple via the nonlinear interaction Hamiltonian. This leads to a new decay channel of the two-polariton state. For the following estimate of the strength of the nonlinearities, we will neglect the influence of this new decay channel, assuming that it is weak. If the two-polariton decay into polaritonic states with other in-plane momenta becomes non-negligible when compared to the single-polariton decay channel (there is no new nonlinear decay channel for single-polariton states), this would lead to non-Hermitian anharmonicities that would also lead to effective nonlinearities with just a single excitation in the system \cite{ben2023non}. 

To estimate when there are strong nonlinearities with just a single excitation in the system, we define single- and two-polariton states that are localized over an area $A = (2\pi)^2/V_{\Delta k_\parallel}$ for each of the modes $i$:
\begin{align}
 \ket{1_{\Delta k_\parallel,i}} &= \frac{1}{\sqrt{V_B V_{\Delta{k_\parallel}}}} \int_{V_{\Delta{k_\parallel}}} \dif^2 k_\parallel P_{\vec{k}_\parallel, i}^\dagger \ket{0} , \\
    \ket{2_{\Delta k_\parallel,i}} &=    \frac{1}{\sqrt{2} V_B V_{\Delta{k_\parallel}}} \int_{V_{\Delta{k_\parallel}}} \dif^2 k_\parallel \int_{V_{\Delta{k_\parallel}}} \dif^2 k_\parallel^\prime P_{\vec{k}_\parallel, i}^\dagger P_{\vec{k}_\parallel^\prime, i}^\dagger\ket{0}.
\end{align}
If $ | \braket{2_{\Delta k_\parallel,i}| \hat{H} |2_{\Delta k_\parallel,i}} - 2 \braket{1_{\Delta k_\parallel,i}| \hat{H} |1_{\Delta k_\parallel,i}}| > \kappa_{2P} $, i.e., if the frequency shift of the two-polariton state due to the nonlinear interaction is larger than the linewidth of the two-polariton state $\kappa_{2P}$ (whose dependence on $\Delta k_\parallel$ and $i$ is not shown explicitly for simplicity), we have a strong nonlinear response with just a single polariton \cite{chang2014quantum,delteil2019towards,munoz2019emergence}. For the case of the BIC considered in the main text, we can do a single-mode approximation for the field with frequency $\omega_{\vec{k}_\parallel,\mathrm{BIC}}$, light-matter coupling $g_{\vec{k}_\parallel,\mathrm{BIC}}$, and decay rate $\kappa_{\vec{k}_\parallel,\mathrm{BIC}}$. We further assume that $V_{\Delta k_\parallel}$ is small with respect to the dispersion of the BIC and that the mode is resonant with the two-level transition such that $\omega_{\vec{k}_\parallel, \mathrm{BIC}} \approx \omega_E$ and $g_{\vec{k}_\parallel,\mathrm{BIC} } \approx g_{\vec{k}_\parallel^\prime,\mathrm{BIC}}$ for all $\vec{k}_\parallel, \vec{k}_\parallel^\prime \in V_{\Delta k_\parallel}$. We than have $V_{\vec{k}_\parallel,1,1}  = V_{\vec{k}_\parallel,1,2} = V_{\vec{k}_\parallel,1,2}  = 1/\sqrt{2}$ and $V_{\vec{k}_\parallel,2,2}  = -1/\sqrt{2}$, such that 
\begin{align} \label{eq:nonlinearshift2}
   | \braket{2_{\Delta k_\parallel,1}| \hat{H} |2_{\Delta k_\parallel,1}} - 2 \braket{1_{\Delta k_\parallel,1}| \hat{H} |1_{\Delta k_\parallel,1}} | =  \frac{V_{\Delta k_\parallel}}{V_\mathrm{B}} g_{\vec{k}_\parallel,\mathrm{BIC}} =  \frac{g_{\vec{k}_\parallel,\mathrm{BIC}}}{N_\mathrm{UC}} > \kappa_{2P}.
\end{align}
Here, $N_\mathrm{UC} = A/V_C$ is the number of unit cells covered by the polaritonic states. At $\vec{k}_{\parallel}=6.4\times10^{-3}\pi/a \hspace{0.08cm}\vec{e}_y$, as considered in Fig.~\ref{fig:BIC_vs_a} below, we have $g_{\vec{k}_\parallel, \mathrm{BIC}} \approx 10^{-4} $\,eV and the decay rate of the two-polariton state reads $\kappa_{2P} =2\kappa_{\vec{k}_\parallel, \mathrm{BIC}} = 1.5 \times 10^{-6}$\,eV, such that Eq.~\eqref{eq:nonlinearshift2} reduces to $ g_{\vec{k}_\parallel, \mathrm{BIC}}/\kappa_{2P} = 66.7>N_\mathrm{UC}$. This inequality is satisfied up to the point where the polaritons span $66.7$ unit cells, which corresponds to an area of $10.7 \,\mu\mathrm{m}^2$ for the square lattice with lattice constant $a=400\,$nm.

\subsection{Driving in the linear regime}

We add a laser driving term $\hat{H}_\mathrm{L} = \Omega (\hat{b}^\dagger_{\vec{k}_\mathrm{L}}+ \hat{b}_{\vec{k}_\mathrm{L}}   )$ (compare Section~\ref{sec:Driving}) to the linear Hamiltonian in the low-excitation regime, such that the Hamiltonian in the co-rotating frame and in the linear approximation at the laser in-plane momentum reads
\begin{align} \label{eqapp:H_linear_driven}
\hat{H}_{\vec{k}_\mathrm{L}, \mathrm{lin}} = (\omega_\mathrm{E}-\omega_\mathrm{L})  \hat{b}_{\vec{k}_\mathrm{L}}^\dagger \hat{b}_{\vec{k}_\mathrm{L}} + \hat{H}_\mathrm{L} + \sum_{ij} (\omega_{\vec{k}_\mathrm{L},ij}  -\omega_\mathrm{L} \delta_{ij}) \hat{a}_{\vec{k}_\mathrm{L},i}^\dagger \hat{a}_{\vec{k}_\mathrm{L},j}  +   \sum_{i} ( g_{ \vec{k}_\mathrm{L},i} \hat{b}_{\vec{k}_\mathrm{L}}^{\dagger}    \hat{a}_{\vec{k}_\mathrm{L},i} + \mathrm{h.c.}) .
\end{align}
This reduces the equations of motion of the emitters and field modes at the laser in-plane momentum to those of two coupled driven lossy harmonic oscillators, for which we can find steady-state solutions analytically \cite{scully1997quantum}. The steady-state solutions for the driven emitter and field operators read $\hat{b}_{\vec{k}_\mathrm{L}, \mathrm{ss}} = \mathcal{b}_{\vec{k}_\mathrm{L}} + \hat{b}_{\vec{k}_\mathrm{L}}$ and $\hat{a}_{\vec{k}_\mathrm{L}, i, \mathrm{ss}} = \mathcal{a}_{\vec{k}_\mathrm{L},i} + \hat{a}_{\vec{k}_\mathrm{L},i}$, respectively, with their coherent amplitudes given by
\begin{align}
\mathcal{A}_{\vec{k}_\mathrm{L}} = -\tens{H}_{\mathrm{eff}}^{-1} \cdot \boldsymbol{\Omega}.
\end{align}
Here, $\mathcal{A}_{\vec{k}_\mathrm{L}} = (\mathcal{a}_{1,\vec{k}_\mathrm{L}},\dots, \mathcal{a}_{N_M,\vec{k}_\mathrm{L}}, \mathcal{b}_{\vec{k}_\mathrm{L}})$, $\boldsymbol{\Omega} = (0,\dots, 0, \Omega)$, and the effective non-Hermitian matrix $\tens{H}_\mathrm{eff}$ reads
\begin{align}
    \tens{H}_\mathrm{eff} = \left( \begin{array}{cc}
      \widetilde{\tens{h}}_{\vec{k}_\mathrm{L}}   & \vec{g}^\ast_{\vec{k}_\mathrm{L}}    \\
   \vec{g}_{\vec{k}_\mathrm{L}}        &       \omega_E -\omega_\mathrm{L}
    \end{array}\right),
\end{align}
with $\widetilde{h}_{\vec{k}_\mathrm{L},ij} = \omega_{\vec{k}_\mathrm{L},ij}-\omega_\mathrm{L}\delta_{ij} -\mi \kappa_{\vec{k}_\mathrm{L},i} \delta_{ij}/2 $. Note that in the linear approximation, all other field and emitter operators with $\vec{k}_\parallel \neq \vec{k}_\mathrm{L}$ remain in the ground state.

In the linear approximation, the field (multiplied with the dipole moment of the emitter so that it has units of energy) seen by the emitter is therefore given by $E_\mathrm{loc} = \Omega + \sum_i g_{\vec{k}_\mathrm{L},i} \mathcal{a}_{\vec{k}_\mathrm{L},i}$, which is shown for the case of the dielectric emitter array in Fig.~3(b) of the main text.

\subsection{Nonlinear interaction in the presence of driving}

To take into account nonlinearities in the driven setup, we consider the nonlinear interaction Hamiltonian in Eq.~\eqref{eqapp:HNL} and also substitute the Holstein-Primakoff expansion in Eq.~\eqref{eqapp:HolsteinPrimakoff_kspace} into the laser driving Hamiltonian $H_\mathrm{L}$. Assuming that the coherent displacement $\mathcal{A}_{\vec{k}_\mathrm{L}}$ of the field and emitter modes at $\vec{k}_\mathrm{L}$ due to the driving field is much stronger than the populations of any other state with $\vec{k}_\parallel \neq \vec{k}_\mathrm{L}$, we find the following two additional nonlinear contributions to the linear Hamiltonian: A squeezing-type interaction
\begin{align} \label{eqapp:Hsqueezing}
 \hat{H}_{\mathrm{sq},\vec{k}_\parallel} = \sum_{i} g_{\vec{k}_\parallel,i} ^\ast\mathcal{b}_{\vec{k}_\mathrm{L}}^2 \hat{a}_{\vec{k}_\parallel,i}^\dagger \hat{b}^\dagger_{\kparb}  + E_\mathrm{loc}\mathcal{b}_{\vec{k}_\mathrm{L}}  \hat{b}_{\vec{k}_\parallel}^\dagger \hat{b}_{\kparb}^\dagger + (\vec{k}_\parallel \leftrightarrow \kparb)  ] + \mathrm{h.c.}, 
\end{align}
and a beam-splitter-type interaction
\begin{align} \label{eqapp:H_beam_splitter}
 \hat{H}_{\mathrm{bs},\vec{k}_\parallel} =  \sum_{i} g_{\vec{k}_\parallel,i} ^\ast |\mathcal{b}_{\vec{k}_\mathrm{L}}|^2 \hat{a}_{\vec{k}_\parallel,i}^\dagger \hat{b}_{\kparb}  + E_\mathrm{loc}\mathcal{b}^\ast_{\vec{k}_\mathrm{L}}  \hat{b}_{\vec{k}_\parallel}^\dagger \hat{b}_{\kparb} + (\vec{k}_\parallel \leftrightarrow \kparb)  ] + \mathrm{h.c.} 
\end{align}
Here, $ (\vec{k}_\parallel \leftrightarrow \kparb) $ denotes adding the preceding terms subject to the replacement $\vec{k}_\parallel \leftrightarrow \kparb$. $\hat{H}_{\mathrm{sq},\vec{k}_\parallel} $ denotes a multimode squeezing Hamiltonian also given in the main text that describes processes in which two excitations of the coherently displaced modes or emitters generate or annihilate pairs of entangled excitations with opposite (but potential finite) in-plane wave-vector $\vec{k}_\parallel$. The beam-splitter Hamiltonian leads to a shift in the polaritonic frequencies, but is of minor importance in our discussion of nonlinear light generation here. 

Figure~4(b) of the main text has been obtained by first finding the steady-state solutions $\mathcal{b}_{\vec{k}_\mathrm{L}}$ and $\mathcal{a}_{\vec{k}_\mathrm{L},i}$ of the equations of motion given by the linear parts of the Hamiltonian at the laser frequency and the additional Lindblad terms, as given by Eq.~\eqref{eq:eom} and Eq.~\eqref{eqapp:H_linear_driven}. Subsequently, $\mathcal{b}_{\vec{k}_\mathrm{L}}$ and $\mathcal{a}_{\vec{k}_\mathrm{L},i}$ are used as input for the non-linear Hamiltonian of modes with $\vec{k}_\parallel \neq \vec{k}_\mathrm{L}$ in Eqs.~\eqref{eqapp:Hsqueezing} and \eqref{eqapp:H_beam_splitter}. The steady-state populations of these modes in the presence of the linear and nonlinear Hamiltonian as well as the Linblad terms were then calculated numerically to obtain $\gamma_{\vec{k}_\parallel}$ given by
\begin{align}
    \gamma_{\vec{k}_\parallel} = \sum_{i,j} ( \kappa_{\vec{k}_\parallel, i} +\kappa_{\overline{\vec{k}}_\parallel, j} )\braket{\hat{a}_{\vec{k}_\parallel, i}^\dagger \hat{a}_{\overline{\vec{k}}_\parallel, j}^\dagger \hat{a}_{\vec{k}_\parallel, i} \hat{a}_{\overline{\vec{k}}_\parallel, j}}_\mathrm{ss}.
\end{align}

\section{Dependence of metasurface resonances on the lattice periodicity}

The coupling of a quantum emitter array to metasurface resonances is strongly dependent on the lattice periodicity. We show this dependence in \autoref{fig:SLR_vs_a} and \autoref{fig:BIC_vs_a} for the plasmonic and dielectric nanoparticle arrays discussed in the main text, respectively. 

For a fixed in-plane wavevector $\left(\vec{k}_{\parallel}=\vec{0}\equiv\Gamma\right.$ for the plasmonic array and $\vec{k}_{\parallel}=6.4\times10^{-3}\pi/a\hspace{0.07cm}\vec{e}_y$ for the dielectric array$\left.\right)$, we display how the reciprocal-space density changes with the lattice constant in \autoref{fig:SLR_vs_a} (a) and \autoref{fig:BIC_vs_a} (a). For the plasmonic array, four different SLRs are visible. Notice that both their frequency and linewidth decrease when increasing the lattice constant. This behavior is a direct consequence of the larger spectral width between the plasmon resonance $\left(\omega_{\mathrm{plasmon}}\sim3.5\hspace{0.08cm}\mathrm{eV}\right)$ and the diffraction orders (DOs) of the lattice when the lattice parameter is increased. Recall that the frequency of DOs scales roughly as $c/a$ \cite{garciadeabajo2007}. Regarding the dielectric array, the BIC behavior is slightly different from the previous case. Its frequency is lower for larger lattice parameters, but its linewidth does not follow a strictly downward trend as in the SLR case. It reaches a minimum around $585$ nm, but it is soon broadened because of the hybridization with the first DO, losing its quasi-Lorentzian shape for a lattice parameter $\sim 800$ nm. A quantitative estimation of the linewidth, $\kappa_c$, and of the emitter-resonance coupling, $g$, are given in \autoref{fig:SLR_vs_a} (b) and \autoref{fig:BIC_vs_a} (b) for the low-energy SLR and the BIC, respectively. The value of $\kappa_c$ for the SLR and the BIC is given by the full width at half maximum (fwhm) of the corresponding resonance. The emitter-resonance coupling $g$ is estimated by mimicking the reciprocal-space spectral density around the SLR and the BIC by a single Lorentzian $J_{\mathrm{lor}}\left(\omega\right)=\frac{g^2}{\pi}\frac{\kappa_c/2}{\left(\omega-\omega_{\mathrm{res}}\right)^2+\left(\kappa_c/2\right)^2}$, where $\omega_{\mathrm{res}}$ stands for the frequency at which the SLR or BIC reciprocal-space spectral density, $J\left(\omega\right)$, is maximum. We thus equate $J_{\mathrm{lor}}\left(\omega_{\mathrm{res}}\right)=J\left(\omega_{\mathrm{res}}\right)$, resulting in the following expression for the emitter-resonance coupling:
\begin{equation}
    g = \sqrt{\frac{\pi}{2}J\left(\omega_{\mathrm{res}}\right)\kappa_c}.
\end{equation}

\begin{figure}[h]
    \centering
    \includegraphics[width=1\linewidth]{SLR_vs_a_horizontal.png}
    \caption{\textit{Surface lattice resonances (SLRs) in a plasmonic array}. (a) Reciprocal-space spectral density $J\left(\Gamma,\omega\right)$ at fixed $\vec{k}_{\parallel}=\vec{0}\equiv\Gamma$ for an array of silver nanoparticles of radius $R=50$ nm as a function of frequency $\omega$ and the lattice constant $a$ of the array. (b) Resonance linewidth $\kappa_c$ and emitter-resonance coupling $g$ for the low-energy SLR as a function of the the lattice constant. (c) Analysis of the strong coupling condition $\mathcal{C}_{\mathrm{SC}}>1$ for the emitter-low-energy-SLR coupling, considering different values for the emitter linewidth $\kappa_e$.}
    \label{fig:SLR_vs_a}
\end{figure}

\begin{figure}[h]
    \centering
    \includegraphics[width=1\linewidth]{BIC_vs_a_horizontal.png}
    \caption{\textit{Bound state in the continuum (BIC) in a dielectric array}. (a) Reciprocal-space spectral density $J\left(\vec{k}_{\parallel},\omega\right)$ at fixed $\vec{k}_{\parallel}=6.4\times10^{-3}\pi/a \hspace{0.08cm}\vec{e}_y$ for an array of silicon nanoparticles of radius $R=100$ nm as a function of frequency $\omega$ and the lattice constant $a$ of the array. (b) Resonance linewidth $\kappa_c$ and emitter-resonance coupling $g$ for the BIC as a function of the the lattice constant. (c) Analysis of the strong coupling condition $\mathcal{C}_{\mathrm{SC}}>1$ for the emitter-BIC coupling, considering different values for the emitter linewidth $\kappa_e$.}
    \label{fig:BIC_vs_a}
\end{figure}

\noindent Note that the aim here is to obtain an approximate value for the linewidth and emitter-resonance coupling that allows us to explore the possibilities of these two platforms to be used as a polaritonic quantum metasurface. An accurate description of the polaritonic dynamics would instead require the use of the few-mode quantization presented in the main text, as especially the SLR resonances are not given by Lorentzian resonances. 

Interestingly, we find that the coupling strength $g\approx 100\,\mu$eV of the BIC is approximately independent of the lattice size, while its decay rate stays below $2\,\mu$eV for lattice constants below $650\,$nm, i.e., as long as the BIC is still far from the first DO. The low-energy SLR reaches similar if not slightly bigger coupling strengths that decrease with the lattice size as its plasmonic character disappears. However, the decay rate decreases even faster with lattice size, which is consistent with recent work showing that the quality factor $Q = \omega_c/\kappa_c$ of SLRs increases when the SLRs are closer to the DO \cite{liang2024local}, as is the case when the lattice size is increased.

Now the onset of strong coupling between the emitter array and the SLR or BIC resonances can be estimated when the condition $\mathcal{C}_{\mathrm{SC}}>1$ is fulfilled, with $\mathcal{C}_{\mathrm{SC}}$ defined as \cite{Torma2015}
\begin{equation}
    \mathcal{C}_{\mathrm{SC}}=\frac{2g}{\sqrt{\kappa_c^2/2+\kappa_e^2/2}}.
\end{equation}
Here, $\kappa_e$ stands for the single-emitter linewidth, which can have different contributions: radiative, dephasing, inhomogeneous broadening, etc. Note that the radiative contribution to this linewidth is already included in $\kappa_c$. We plot $\mathcal{C}_{\mathrm{SC}}$ in \autoref{fig:SLR_vs_a} (c) and \autoref{fig:BIC_vs_a} (c) for different values of the emitter linewidth. This estimation predicts the onset of the strong coupling in the plasmonic array for $a>615$ nm, while the strong coupling survives in the dielectric array with a remarkably larger value of $\mathcal{C}_{\mathrm{SC}}$ for all the spectral width in which the BIC is not hybridized with the first DO. We thus find that only for large lattice sizes can the SLRs reach the strong coupling regime because of the much higher decay rates of the SLRs compared to the BIC.

This general comparison is in-line with state-of-the-art experiments that reach higher $Q$-factors ($Q \equiv \omega_c/\kappa_c$) for quasi BICs ($Q= 18500$ \cite{liu2019high}, corresponding to $\kappa_c = 97 \,\mu$eV for $\omega_c = 1.8\,$eV) than for SLRs ($Q = 2340$ \cite{bin2021ultra}, corresponding to $\kappa_c = 1$meV for $\omega_c = 2.5\,$eV). So also these experimentally determined values of $Q$ and the corresponding values of the decay rate $\kappa_c$, together with the coupling strengths determined in our manuscript, put the emitter-BIC coupling in the strong coupling regime. The slightly larger estimate of $\kappa_c$ obtained from the experimentally determined $Q$-factor, for the BIC, compared to our simulations, is a result of finite size effects and imperfections in the experimentally realized nanoparticle array, as well as of how much the symmetry is broken in the experiment and in the simulation (by choosing a finite value for $k_\parallel$), which turns the dark BIC (with $\kappa_c \approx 0$) into a quasi BIC with finite $\kappa_c$ that can be probed experimentally. 

For the emitter-SLR coupling, on the other hand, we find that for the setup analyzed here strong coupling is only achieved for lattice size $a >615$\,nm, in which case the $Q$-factor of the array that is needed to reach strong coupling is $Q \approx 3300$, which is slightly bigger than the experimentally achieved record-high value of $Q = 2340$ \cite{bin2021ultra}. In the manuscript, we therefore show results with a lattice constant of $a = 600\,$nm for the plasmonic nanoparticle array, which corresponds to a quality factor at the $\Gamma$ point of $Q= 2000$, which is similar to that achieved experimentally.

\section{Quantum emitter arrays in free space}

A quantum emitter array in free space can be understood as a particular limiting case of the platform introduced in this work when no metasurface is included. In this configuration, the resonances to which the emitter array can couple are the diffraction orders (DOs) of the empty lattice. The developed formalism still applies here, as we discuss in the following. 

We compare the performance of a free-space array of quantum emitters, when coupled to a DO, to the BIC-based polaritonic quantum metasurface discussed in the main text. We consider a free-space array of lattice constant $a=400$ nm with a x-oriented emitter in the unit cell at $\vec{r} = \boldsymbol{0}$, matching in this way the emitter orientation and lattice configuration for the BIC-based polaritonic quantum metasurface. Unlike general metasurfaces, analytical insight on the Green tensor can be provided for a free-space array. Inserting the $(2+1)$ Weyl decomposed free-space Green tensor \cite{Buhmann2012a}
\begin{align}
   G_{xx}(\vec{r},\vec{r}^\prime,\omega) = \frac{\mi}{8\pi^2}\int \dif^2 k_\parallel \frac{\me^{\mi \vec{k}_\parallel (\vec{r}-\vec{r}^\prime)} }{\sqrt{k^2-k_\parallel^2}} \left( 1- \frac{k_x^2}{k^2}\right),
\end{align}
into Eq.~(1) of the main text yields an analytical expression for the reciprocal-space spectral density of a quantum emitter array in free space: 
\begin{align}
    J_{\mathrm{0},xx} (\omega, \vec{k}_\parallel)= \frac{\mu_0 \omega^2 d^2V_\mathrm{B}}{2\pi \hbar } \sum_{\mathbf{K}} \mathrm{Re}\left[\frac{1-\frac{(k_x+K_x)^2}{k_\parallel^2}}{\sqrt{\frac{\omega^2}{c^2}   - (\mathbf{k}_\parallel + \mathbf{K})^2} }  \right].
\end{align}
As we can see, for an ideal infinite quantum emitter array, $J_\mathrm{0}$ discontinuously jumps from a finite value to infinity when approaching the DO frequency, $\omega_\mathrm{DO}$, from below and falls as $1/\sqrt{\omega}$ for $\omega > \omega_\mathrm{DO}$. The DOs are therefore highly non-Lorentzian resonances that, due to the discontinuity at $\omega_\mathrm{DO}$, can only be approximated by a large set of coupled quantized modes. The divergency at the DO is a result of the idealized scenario of a perfect infinite array in free space, and it becomes finite in the presence of an optical environment (such as a metasurface, as studied in this work), for finite quantum emitter arrays, losses in the background medium or for imperfect arrays.

Although this divergency makes the exact analysis challenging, we can still estimate the coupling strength $g(\vec{k}_\parallel)$ of a quantum emitter array in free space to the DOs in a single-mode approximation:
\begin{align}
  g^2(\vec{k}_\parallel) \approx  \int_{\omega_\mathrm{DO}-\Delta \omega/2}^{\omega_\mathrm{DO}+\Delta \omega/2} \dif \omega J(\omega) ,
\end{align}
as follows from Eq.~(3) of the main text. The width of the resonance $\Delta \omega$ should be chosen so that it covers the resonant peak. We find 
\begin{align}
  \hbar  g \approx  =\sqrt{ \frac{\hbar \mu_0 \omega^2 d^2V_B}{2\pi  } \mathrm{arccosh}\left[ 1 + \Delta \omega\right]} \approx 15\, \mu \mathrm{eV},
\end{align}
for $\Delta \omega = 5\, \mu\mathrm{eV}$, dipole strength of the emitters $d = 10\,$D, and a square lattice with lattice constant $a=400\,$nm and $\omega_\mathrm{DO} = 2.915\,$eV. This coupling strength is at least an order of magnitude smaller when compared to the coupling strengths achieved by the bound state in the continuum and the surface lattice resonances considered in this work, see section SIV of the Supplemental Material.

\end{document}